\documentclass{article}
\usepackage{spconf,amsmath,graphicx,hyperref}
\usepackage{tikz}
\usetikzlibrary{arrows.meta,positioning,shapes.geometric,fit,backgrounds}
\usepackage{subcaption}
\usepackage{caption}
\usepackage{amssymb}   % for arrow symbols
\usetikzlibrary{positioning}
\usetikzlibrary{decorations.pathreplacing,positioning}
\usepackage{booktabs,multirow,tabularx,xcolor,colortbl}
\definecolor{headorange}{RGB}{230,115,50}
\definecolor{groupblue}{RGB}{206,222,246}
\definecolor{groupgreen}{RGB}{208,233,208}
\usepackage{float}
\usepackage{multirow}
\usepackage{bigdelim} % for curly brackets

\title{BAmbiDrop: Array-Agnostic Speech Enhancement with Zero-Shot binaural Reproduction using Multi-center Ambisonics Encoding}
\name{Michael Tatarjitzky$^{1}$, Vladimir Tourbabin$^{2}$, Boaz Rafaely$^{1}$}
\address{$^{1}$School of Electrical and Computer Engineering, Ben Gurion University of the Negev, Beer-Sheva, Israel. \\
$^{2}$Reality Labs Research at Meta, Redmond, WA, USA.}
\begin{document}
\ninept
\topmargin=0mm 
\maketitle
\copyrightnotice{\parbox{\textwidth}{\footnotesize
\copyright~2027 IEEE. Personal use of this material is permitted. Permission from IEEE must be obtained for all other uses, in any current or future media, including reprinting/republishing this material for advertising or promotional purposes, creating new collective works, for resale or redistribution to servers or lists, or reuse of any copyrighted component of this work in other works.}}

\begin{abstract}
Array-agnostic enhancement networks generalize across microphone geometries but produce a single channel, while binaural enhancement methods preserve interaural cues only for one fixed device geometry. We propose BAmbiDrop, a framework that extends array-agnostic enhancement to binaural reproduction without retraining. Exploiting the fact that the Ambisonics origin is an arbitrary point in space, a frozen single-channel AmbiDrop network estimates a complex mask at multiple centers across the array; these masks are fused by a complex mean into a single shared mask, applied to the microphone signals and rendered by fixed BSM filters. Across 21 free-field and rigid-sphere arrays, the fused mask improves enhancement and interaural cue reproduction simultaneously, whereas masking each microphone independently degrades the latter without a consistent gain in the former.
\end{abstract}
\begin{keywords}
Binaural reproduction, speech enhancement, Ambisonics, array-agnostic modeling, spatial cue preservation
\end{keywords}

\section{Introduction}
\label{sec:intro}

Capturing an acoustic scene with a microphone array and reproducing it for a listener is a central problem in spatial audio, underlying immersive telepresence, augmented reality, hearing assistance and teleconferencing \cite{rafaely2022spatial}. In binaural reproduction, the recording is rendered to the two ears of a listener, such that the interaural cues supporting localization and externalization are preserved \cite{rafaely2022spatial}. Real scenes may contain competing talkers, noise and reverberation, so the recording must also be enhanced for the target talker to be intelligible \cite{osaughnessy2024speech}. Achieving high performance on both operations may therefore impose a challenge \cite{rafaely2022spatial}.
%Both operations act on the same signals, produced by the specific array of the capturing device.

%The geometry of these arrays, however, varies from one device to the next \cite{donley2021easycom}. 

Systems that address both objectives differ mainly in the domain in which the noise reduction is applied. One line operates in the Ambisonics domain, where the enhanced signal remains a sound-field representation that can still be rendered binaurally, by masking designed for binaural reproduction \cite{lugasi2020masking}, direction- and reverberation-preserving noise reduction \cite{herzog2020direction}, or deep Ambisonics-to-Ambisonics networks \cite{herzog2023ambisep, lugasi2024multichannel}. A second line operates at the ears, where the trade-off between noise reduction and cue preservation is well documented: beamformers for binaural hearing devices constrain the interaural cues of the residual signal explicitly \cite{cornelis2010theoretical, marquardt2015interaural}, and learning-based systems produce a two-channel output under cue-preservation loss terms \cite{tokala2024binaural}. A third line enhances the microphone signals themselves, estimating one output per microphone so that a renderer may follow \cite{pandey2022tparn, lugasi2022spatial}, an architecture recently motivated by binaural reproduction \cite{deppisch2026direction}. Although these methods were found useful, the learning-based ones are typically preferable over classical approaches due to their improved speech enhancement \cite{huang2025advances}. 
%optimized for a single microphone array, and the Ambisonics-domain ones presume an input of sufficient order, which in practice requires a spherical array \cite{rafaely2015fundamentals}, whereas the classical, signal-independent designs accommodate a wider range of arrays but achieve lower enhancement performance. A network 
However, learning-based neural networks are usually optimized for one microphone array layout and generalize poorly to others \cite{tatarjitzky2026ambidrop}, and although they can be retrained per device, this requires substantial data which may not always be available. This shortcoming has motivated array-agnostic processing approaches.

Array-agnostic enhancement has been approached by architectures that remain invariant to the number and the placement of the microphones, employing transform-average-concatenate layers \cite{luo2020end} or attention across channels \cite{yoshioka2022vararray}, at the cost of large multi-geometry training sets. Recently, we proposed AmbiDrop \cite{tatarjitzky2026ambidrop, ambidropIEEE}, which instead trains on ideal Ambisonics with channel-wise dropout and encodes an arbitrary array by Ambisonics Signal Matching (ASM) \cite{gayer2024ambisonics} at inference, so that no array geometry is observed during training. These methods, however, produce a single channel, and the spatial attributes of their output are not preserved. The closest exception maps array signals to a binaural output through a network trained with a geometry-invariant coherence feature \cite{hsu2023telepresence}, but its renderer is itself learned from data of one array family, and its invariance is demonstrated on a small set of compact configurations. In summary, the development and study of methods for array-agnostic speech enhancement that also deliver binaural audio remains a gap in current literature. 

%Two gaps therefore remain: methods that preserve the interaural cues are tied to the array they were designed or trained for, whereas array-agnostic methods are single-channel and leave open how the enhanced signal reaches the ears. 

We propose BAmbiDrop to bridge this gap. The proposed method applies the mask of a trained AmbiDrop network \cite{tatarjitzky2026ambidrop} by re-positioning the Ambisonics coordinate-system origin to align with each of the array microphones, then applying Binaural Signal Matching (BSM) filters \cite{madmoni2020beamforming} to the enhanced microphone signals to produce binaural signals.
%, matching the head-related transfer function (HRTF) at each ear from the array steering matrix alone and therefore applicable to an arbitrary geometry. 
The method exploits the 
%arbitrariness of the Ambisonics origin
generalization of the spherical harmonics representation: relocation can be applied to any microphone position using the same trained mask without loss of performance. 
%Since a mask estimated at one center describes the pressure at that point alone, averaging the masks obtained at centers spanning the aperture reduces the error of the applied mask, while a single shared mask multiplies every channel by the same complex value and thereby preserves the inter-channel relations on which BSM depends. 
We further show that a common mask shared between microphones provides a more accurate binaural signal as it reduces inter-channel variability compared to microphone-dependent masks.
% also offering a more computationally efficient solution.  
%
The contributions of this work are as follows: (i) the development of an array-agnostic speech enhancement framework with an accurate and efficient projection to binaural signals, (ii) once trained, the network can be applied to any array without re-training, (iii) the proposed method has been 
%are a binaural output from an arbitrary array with no component retrained, a fusion rule that improves enhancement and interaural cue reproduction together rather than trading them against each other, 
evaluated for both speech enhancement and binaural reproduction, showing improved performance compared to baselines.

\section{Background}
\label{sec:background}
 
This section presents the array signal model, Ambisonics encoding by ASM, binaural reproduction by BSM, and the AmbiDrop enhancement model. The spherical coordinate system $(r,\theta,\phi)$ denotes radius, elevation and azimuth, and the wavenumber is $k = \frac{2\pi}{c}f$, with $c$ the speed of sound and $f$ the frequency.
 
\subsection{Array Signal Model}
\label{ssec:model}
 
Consider an arbitrary array of $D$ omnidirectional microphones positioned at $(r_d,\theta_d,\phi_d)$, $\forall\, 1 \le d \le D$, in a sound field composed of
$Q$ plane waves which represent the direct sound from far-field sources
together with their reflections from the enclosure boundaries. The plane-wave arrival directions are denoted by $\mathbf{u}_q=(\theta_q,\phi_q)$,
$\forall\, 1 \le q \le Q$. 
The array steering matrix $\mathbf{V}(k)$ of dimensions $D \times Q$, with
element $[\mathbf{V}(k)]_{\mathit{d,q}}$ represents the transfer function between the $q$-th
plane wave and the $d$-th microphone. With this model, the microphone signals can be formulated as:
\begin{equation}
\mathbf{x}(k) = \mathbf{V}(k)\mathbf{s}(k) + \mathbf{n}(k),
\label{eq:model}
\end{equation}
where $\mathbf{s}(k)$ holds the plane-wave amplitudes at the origin and
$\mathbf{n}(k)$ is additive sensor noise, assumed spatially white and
uncorrelated with the source signals. 
%Throughout, the free-field steering convention $e^{-jk\mathbf{u}_q\cdot\mathbf{r}_m}$ is adopted.
 
\subsection{Ambisonics Encoding and ASM}
\label{ssec:asm}
 
The Ambisonics signals up to order $N_{\mathit{a}}$ are obtained from the plane-wave
amplitudes as $\mathbf{a}(k) = \mathbf{Y}^{H}_{\mathit{\Omega_Q}}\mathbf{s}(k)$ \cite{rafaely2015fundamentals}, where $\mathbf{a}(k)$ stacks the Ambisonics signals $a_{\mathit{nm}}(k)$ and $\mathbf{Y}_{\mathit{\Omega_Q}}$ is the SH matrix of dimensions $Q \times (N_{\mathit{a}}+1)^2$ whose columns $\mathbf{y}_{\mathit{nm}}$ contain the SH basis functions of order $0 \leq n \leq N_{\mathit{a}}$ and degree $-n \leq m \leq n$ evaluated at the $Q$ plane-wave directions. This representation depends on the sound field alone and not on the array that observes it.
 
ASM \cite{gayer2024ambisonics} estimates these signals from an arbitrary array by a linear mapping of the microphone signals,
\begin{equation}
\hat{a}_{\mathit{nm}}(k) = {[\mathbf{c}_{\mathit{nm}}^{ASM}(k)]}^H\mathbf{x}(k),
\label{eq:asm}
\end{equation}
with the coefficients $\mathbf{c}(k)$ designed to minimize the NMSE between the
estimated and the true Ambisonics signals. Under a diffuse field with source
covariance $\sigma_{\mathit{s}}^2\mathbf{I}$ and spatially white noise of power $\sigma_{\mathit{n}}^2$, the minimizer is:
\begin{equation}
\mathbf{c}_{\mathit{nm}}^{ASM}(k) =
\left( \mathbf{V}(k)\mathbf{V}^H(k) + \frac{\sigma_{\mathit{n}}^2}{\sigma_{\mathit{s}}^2} \mathbf{I} \right)^{-1}
\mathbf{V}(k) \mathbf{y}_{\mathit{nm}},
\label{eq:c_opt}
\end{equation}
which requires $(N_{\mathit{a}}+1)^2 \le D$ for all channels to be encoded \cite{rafaely2015fundamentals}. The accuracy of individual channels nonetheless depends on the array geometry, and so some channels
may be poorly estimated~\cite{gayer2024ambisonics}.
 
\subsection{Binaural Reproduction With BSM}
\label{ssec:bsm}

BSM~\cite{madmoni2020beamforming} reconstructs the binaural signal at a listener's ears from signals recorded by an arbitrary microphone array, sharing the same linear structure as ASM (Eq.~\ref{eq:asm}).  The binaural signals $\hat{p}_{\mathit{l,r}}(k) = {[\mathbf{c}_{\mathit{l,r}}^{BSM}(k)]}^{H}\mathbf{x}(k)$, combined into vector $\mathbf{p}(k) = [p_{\mathit{l}}(k), p_{\mathit{r}}(k)]^{T}$, are estimated by the filter $\mathbf{c}_{\mathit{l,r}}^{BSM}(k)$. This filter minimizes the NMSE to the true binaural signals, and has a form similar to Eq.~(\ref{eq:c_opt}).
%with the SH target $\mathbf{y}_{\mathit{nm}}$.
%replaced by the HRTF $\mathbf{h}_{\mathit{l,r}}(k)$ of the corresponding ear~\cite{madmoni2020beamforming}. 
Above a crossover frequency the least-squares fit is replaced by a magnitude least-squares minimization~\cite{madmoni2020beamforming}, which relaxes the phase constraint where interaural phase is perceptually less important.
% Because both designs invert the same regularized array covariance, the two encoders share the geometry dependence discussed above. Moreover, since the filters are linear and shared across the channels, the interaural cues that BSM reproduces are determined by the relations between the microphone signals rather than by any single channel, a property that the method of Section~\ref{sec:method} is designed to preserve.

\subsection{Speech Enhancement With AmbiDrop}
\label{ssec:ambidrop}

AmbiDrop~\cite{ambidropIEEE} performs the task of speech enhancement  in the Ambisonics domain, using an FT-JNF network~\cite{tesch2022insights} trained exclusively on ideal Ambisonics signals of order $N_{\mathit{a}} = 2$, with channel-wise dropout applied during training. Since these signals depend on the sound field alone, as noted in Sec.~\ref{ssec:asm}, the network training is independent of any microphone array, while the dropout improves robustness to the channel-dependent encoding errors that ASM introduces at inference.

At inference, the microphone signals of an arbitrary array are encoded by ASM, transformed to the STFT domain, and processed by the network, which estimates a complex mask $M(t,f)$ at each time-frequency bin $(t,f)$. The mask is applied to the omnidirectional channel,
\begin{equation}
\tilde{a}_{00}(t,f) = M(t,f)\,\hat{a}_{00}(t,f),
\label{eq:mask}
\end{equation}
yielding a single-channel enhanced signal. The model used here differs from the published one in the training reference alone, which is the reverberant target speaker rather than its direct-path component: reverberation carries the externalization, perceived distance and envelopment cues a binaural renderer must reproduce \cite{catic2015reverberation, bronkhorst1999auditory, bradley1995objective}, a well-documented consideration in spatial audio reproduction.

\section{Proposed Method}
\label{sec:method}
 
\begin{figure*}[t]
\centering
\includegraphics[width=\textwidth]{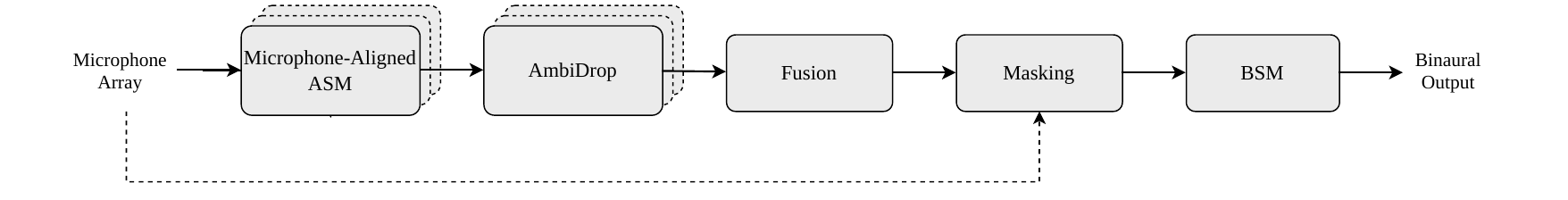}
\caption{The BAmbiDrop pipeline (Sec.~\ref{sec:method}). The microphone array feeds $D$ microphone-aligned ASM encoders, one per microphone position $\mathbf{r}_{\mathit{d}}$ (Sec.~\ref{ssec:bambidrop}); the same frozen AmbiDrop network (Sec.~\ref{ssec:bambidrop}) is evaluated once per microphone, giving one mask $M_{\mathit{d}}(t,f)$ each, and Fusion (Sec.~\ref{ssec:bambidrop}) averages them into a single shared mask $\bar{M}(t,f)$. Masking (Sec.~\ref{ssec:bambidrop}) applies that mask to the unprocessed microphone signals, and BSM (Sec.~\ref{ssec:bambidrop}) renders the result to the binaural signal. Dashed outlines mark repeated instances of the same block, one per microphone.}
\label{fig:pipeline}
\end{figure*}

We propose BAmbiDrop, a binaural extension of the single-channel-output AmbiDrop framework, in which speech enhancement is applied separately to each microphone by repositioning the center of the Ambisonics coordinate system for each microphone across the array, and fusing the microphone-dependent masks into a single shared mask prior to binaural rendering. The proposed method is outlined in Fig.~\ref{fig:pipeline}, and detailed below.

%\subsection{Microphone-Aligned ASM}
\subsection{BAmbiDrop}
\label{ssec:bambidrop}

\textbf{Microphone-Aligned ASM:} The Ambisonics channel for which AmbiDrop applies masking as in Eq. (\ref{eq:mask}) is $a_{00}$ which represent the sound pressure at the origin of the Ambisonics coordinate system \cite{rafaely2015fundamentals}. Because the approach of the proposed method is based on applying the mask computed for channel $a_{00}$ to an array microphone, it is hypothesized that this approach will work better when the microphone is aligned with the origin of the Ambisonics coordinate system. For this reason, the Microphone-Aligned ASM block aligns the origin of the coordinate system with one array microphone before applying ASM and the AmbiDrop mask to that microphone. In the following, this alignment process is presented in detail.

The main effect of this origin alignment is on the steering matrix $\mathbf{V}(k)$. Because it represents the transfer function from a plane-wave of an amplitude defined at the origin to the output of the array microphones, aligning the origin of the coordinate system simply leads to a shift in the plane-wave amplitude, as also exploited in \cite{BenHur2019EarAlignedHRTF} for HRTF alignment. The elements of the aligned steering matrix can therefore be formulated as:
\begin{equation}
[\mathbf{V}^{\boldsymbol{\delta}}(k)]_{\mathit{d,q}} = [\mathbf{V}(k)]_{\mathit{d,q}}\,\exp\!\big(jk\,\mathbf{u}_{\mathit{q}}\cdot\boldsymbol{\delta}\big).
\label{eq:vshift}
\end{equation}
where $\mathbf{u}_{\mathit{q}}$ denotes the direction of the $q$-th plane-wave and $\boldsymbol{\delta}$ denotes the displacement of the origin relative to the origin defined for $\mathbf{V}(k)$. Substituting Eq. (\ref{eq:vshift}) in Eq. (\ref{eq:c_opt}) leads to
\begin{equation}
\mathbf{c}_{\mathit{nm}}^{\boldsymbol{\delta}}(k) =
\left( \mathbf{V}(k)\mathbf{V}^H(k) + \frac{\sigma_{\mathit{n}}^2}{\sigma_{\mathit{s}}^2} \mathbf{I} \right)^{-1}
\mathbf{V}^{\boldsymbol{\delta}}(k)\, \mathbf{y}_{\mathit{nm}},
\label{eq:c_shift}
\end{equation}
where it is noted that $\mathbf{V}^{\boldsymbol{\delta}}(k) [\mathbf{V}^{\boldsymbol{\delta}}(k)]^H = \mathbf{V}(k)[\mathbf{V}(k)]^H$ due to the cancellation of the exponential function.

% \ref{eq:c_opt}) with $\mathbf{V}^{\boldsymbol{\delta}}(k)$ in place of $\mathbf{V}(k)$, and the SH target $\mathbf{y}_{\mathit{nm}}$ unchanged,
% \begin{equation}
% \mathbf{c}_{\mathit{nm}}^{\boldsymbol{\delta}}(k) =
% \left( \mathbf{V}^{\boldsymbol{\delta}}(k){\mathbf{V}^{\boldsymbol{\delta}}}^{H}(k) + \frac{\sigma_{\mathit{n}}^2}{\sigma_{\mathit{s}}^2} \mathbf{I} \right)^{-1}
% \mathbf{V}^{\boldsymbol{\delta}}(k)\, \mathbf{y}_{\mathit{nm}},
% \label{eq:c_shift}
% \end{equation}
% gives $\hat{a}_{\mathit{nm}}^{\boldsymbol{\delta}}(k) = [\mathbf{c}_{\mathit{nm}}^{\boldsymbol{\delta}}(k)]^{H}\mathbf{x}(k)$: the Ambisonics channels of (\ref{eq:asm}), re-centered to the new origin, estimated without re-measuring the array. Since the sources are far-field plane waves, their directions $\mathbf{u}_{\mathit{q}}$ are invariant under the translation, so no parallax is introduced and the target speaker stays at azimuth $0$ regardless of $\boldsymbol{\delta}$; the relation is exact for free-field arrays, and an approximation for arrays on a rigid sphere, where scattering is not captured by this construction alone - both types are evaluated in Section~\ref{sec:experiment}.

BAmbiDrop sets $\boldsymbol{\delta} = \mathbf{r}_{\mathit{d}}$, the position of the $d$-th microphone of Sec.~\ref{ssec:model}, for every $1 \le d \le D$: the microphone signals are re-encoded using ASM $D$ times, once per microphone position, giving $D$ sets of microphone-aligned Ambisonics channels, transformed to the STFT domain for further processing.
%the frozen network of Section~\ref{ssec:infer_ambidrop}.

%\subsection{AmbiDrop}
%\label{ssec:infer_ambidrop}

\textbf{AmbiDrop:} The network of Sec.~\ref{ssec:ambidrop}, trained once on ideal Ambisonics, independent of any array, is frozen (no further training or adaptation), and evaluated $D$ times - once on the microphone-aligned channels of each microphone $d$. Each evaluation returns one complex mask, $M_{\mathit{d}}(t,f)$, the instance of Eq. (\ref{eq:mask}) $M(t,f)$ at that microphone's origin. 
%Weights are shared across all $D$ evaluations, so no component is retrained and no array-specific parameter is introduced.

%\subsection{Fusion}
%\label{ssec:fusion}

\textbf{Fusion:} The AmbiDrop stage as %in Sec. \ref{ssec:infer_ambidrop} 
above estimated a total of $D$ masks, one for each microphone. In this fusion stage, these masks are averaged to form a single mask:
%, A mask estimated at a single origin describes the enhancement of the pressure at that point alone, and since the $D$ microphones of Section~\ref{ssec:model} are distributed around the array, no single center's mask fits every one of them equally well. Fusion instead averages the $D$ per-microphone masks of Section~\ref{ssec:infer_ambidrop} by the unweighted complex mean
\begin{equation}
\bar{M}(t,f) = \frac{1}{D}\sum_{d=1}^{D} M_{\mathit{d}}(t,f).
\label{eq:fusion}
\end{equation}
%into the single shared mask $\bar{M}(t,f)$ that Section~\ref{ssec:masking} applies. Averaging the $D$ per-center estimates reduces the residual of the applied mask relative to any single center, while a single shared mask, common to every channel, 
It is expected that such a mask common to all microphones will reduce inter-channel distortion and improve the binaural reproduction quality that follows. Nevertheless, this may come at the expense of speech enhancement performance. The variations described in Sec. \ref{ssec:altconfig} are designed to study this tradeoff. 

% \subsection{Masking}
% \label{ssec:masking}

%Unlike (\ref{eq:mask}), which applies the mask to the encoded channel $\hat{a}_{00}(t,f)$, BAmbiDrop 
\textbf{Masking:} In this stage, the common mask is applied directly to each of the unprocessed microphone signals,
\begin{equation}
\tilde{X}_{\mathit{d}}(t,f) = \bar{M}(t,f)\,X_{\mathit{d}}(t,f), \quad 1 \le d \le D,
\label{eq:masking}
\end{equation}
leading to speech-enhanced microphone signals ready for binaural reproduction. %bypassing the microphone-aligned encoding of Section~\ref{ssec:bambidrop} entirely, as shown in Fig.~\ref{fig:pipeline}. Every channel is multiplied by the same $\bar{M}(t,f)$, so this step alone determines whether the inter-channel relations Section~\ref{ssec:bsm_render} relies on survive.

% \subsection{BSM}
% \label{ssec:bsm_render}

\textbf{BSM:} The masked signals $\tilde{X}_{\mathit{d}}(t,f)$ are decoded to binaural signals by the fixed BSM filters, detailed in Sec.~\ref{ssec:bsm}. Note that this BSM filter is common to each variation detailed next in Sec.~\ref{ssec:altconfig}.
%, any difference between them is attributable to the mask alone.

\subsection{Variations of the Proposed Method}
\label{ssec:altconfig}

In this paper the proposed method is studied alongside several variations as detailed next.\\
\noindent \textbf{Fusion:} This is the main method, as detailed in Sec. \ref{ssec:bambidrop}.\\
%Three further configurations reuse the same construction with a different choice of origin. 
\textbf{Center:} sets $\boldsymbol{\delta} = \mathbf{0}$ in Eq. (\ref{eq:vshift}), as used originally in \cite{tatarjitzky2026ambidrop} and detailed in Sec.~\ref{ssec:asm}, and applies this single mask to each microphone via Eq.  (\ref{eq:masking}).\\
%\textbf{Ref-Mic:} sets $\boldsymbol{\delta} = \mathbf{r}_0$ for a specific reference microphone denoted $d=0$, chosen as the one closest to the target speaker (see Sec. \ref{sec:experiment}). Note that this single reference mask is applied to all $D$ microphones.\\ 
%at azimuth $0$ - one instance of Section~\ref{ssec:bambidrop}'s construction rather than all $D$ - and likewise applies that single mask to every channel; both are coherence-safe, since a single mask is still common to every channel, but neither benefits from the residual-averaging of Fusion (\ref{eq:fusion}). 
%
\textbf{Per-Mic:} this variation skips the Fusion stage, such that the $D$ masks $M_{\mathit{d}}(t,f)$ are applied individually to each microphone, $\tilde{X}_{\mathit{d}}(t,f) = M_{\mathit{d}}(t,f)X_{\mathit{d}}(t,f)$, therefore not sharing a common mask. 
%Since the $D$ channels are then multiplied by $D$ different complex gains rather than one shared value, 
This configuration may not preserve the inter-channel relations,
%on which Section~\ref{ssec:bsm_render} depends, as Section~\ref{ssec:results} shows.
but may improve speech enhancement due to the microphone-specific masks.

\subsection{Baselines}
\label{ssec:baselines}

Two methods are used as baselines for comparing to the proposed method and its variants.
%The baselines assess not only the array-agnostic design of Secs.~\ref{ssec:bambidrop}-\ref{ssec:masking}, but also a more straightforward MIMO implementation. 
Both use the same FT-JNF architecture as the proposed method, but trained directly on the $D$ microphone signals,
%rather than on the Ambisonics-encoded channels used by AmbiDrop~\cite{ambidropIEEE}, 
extended to a $D$-mask MIMO output, one mask per microphone, in two different variations, as detailed below. Both baselines are trained on a training set of various array geometries.\\
\textbf{MIMO FT-JNF} applies a separate mask to each channel, the baseline analogue of \emph{Per-Mic}.\\
\textbf{MIMO FT-JNF (Fusion)} averages the $D$ masks by a complex mean 
%inside the training graph, 
before applying them, which is the baseline analogue of \emph{Fusion}.

\section{Experiment}
\label{sec:experiment}

This experiment evaluates the performance of the proposed method and its variations 
%target-speaker extraction from a multi-talker mixture, and the binaural reproduction of the enhanced signal. The variations of Sec.~\ref{ssec:altconfig} are compared 
against two baselines and across various array configurations and acoustic conditions.

% \begin{table*}[t]
% \centering
% \begin{tabular}{lccccccc}
% \hline
% Method
% & Enh $\Delta$ dB$\uparrow$
% & Dir-pres dB$\uparrow$
% & C50 err dB$\downarrow$
% & Noise red. dB$\uparrow$
% & ITD $\mu$s$\downarrow$
% & ILDcov dB$\downarrow$
% & IC $\downarrow$ \\
% \hline
% \textit{BSM Clean}
% & 5.8 / 9.2
% & -0.6 / 3.1
% & 0.5 / 0.0
% & 0.0 / 0.0
% & 64 / 28
% & 2.2 / 2.0
% & 0.19 / 0.09 \\

% \hline

% Center
% & 2.7 / 6.1
% & -1.6 / 1.7
% & 1.8 / 1.3
% & 2.3 / 3.5
% & 101 / 59
% & \textbf{1.9} / 2.2& 0.32 / 0.20 \\

% % Ref-Mic
% % & 2.5 / 5.6
% % & -1.5 / 1.6
% % & 1.8 / 1.4
% % & 2.1 / 3.0
% % & 98 / 54
% % & 2.0 / \textbf{2.2}
% % & 0.33 / 0.20 \\

% Per-Mic
% & -0.0 / 5.4
% & -6.8 / 0.0
% & \textbf{0.9} / \textbf{1.1}
% & -2.2 / 3.2
% & 137 / 77
% & 2.9 / 4.8
% & 0.34 / 0.18 \\

% Fusion
% & \textbf{3.2} / \textbf{6.9}
% & \textbf{-1.1} / \textbf{2.4}
% & 1.8 / 1.3
% & \textbf{3.3} / \textbf{5.0}
% & \textbf{96} / \textbf{53}
% & 2.0 / 2.3& \textbf{0.31} / \textbf{0.17} \\
% \hline
% \end{tabular}
% \caption{Mean performance of each mask configuration, reported as free-field / rigid-sphere arrays, with metrics defined in Sec.~\ref{ssec:methodology}. The true noisy ears measure $-5.4\,$dB SI-SDR relative to the true reverberant ears, the floor that Enh $\Delta$ improves upon. The BSM Clean row passes each metric's own clean, noise-free target through the fixed BSM filters with no mask (the best any masking method could reach) and is a reference, not ranked; the best value of each column is in bold.}
% \label{tab:results}
% \end{table*}

\begin{table*}[t]
\centering
\begin{tabular}{lccccccc}
\hline
Method
& SI-SDR dB$\uparrow$
& PESQ $\uparrow$
& STOI $\uparrow$
& $\Delta$IC $\downarrow$
& $\Delta$C50 dB$\downarrow$
& $\Delta$ILD dB$\downarrow$
& SI-SDR-Dir dB$\uparrow$ \\
\hline
\textit{BSM Noisy}
& -8.80
& 1.23
& 0.525
& 0.25
& 0.00
& 5.07
& -11.98 \\

\textit{BSM Clean}
& 2.25
& 3.69
& 0.945
& 0.14
& 0.22
& 1.44
& 1.34 \\

\hline

Center
& -0.87
& 1.54
& 0.756
& 0.26
& 1.54
& 4.77
& 0.13 \\

% Ref-Mic

Per-Mic
& -2.51
& 1.57
& 0.753
& 0.26
& \textbf{1.03}
& 5.07
& -3.23 \\

Fusion
& \textbf{-0.30}
& \textbf{1.62}
& \textbf{0.785}
& \textbf{0.23}
& 1.54
& \textbf{4.71}
& \textbf{0.69} \\

MIMO FT-JNF
& -6.13
& 1.46
& 0.650
& 0.29
& 1.12
& 4.80
& -4.01 \\

MIMO FT-JNF (Fusion)
& -7.16
& 1.45
& 0.634
& 0.28
& 1.18
& 4.72
& -4.68 \\
\hline
\end{tabular}
\caption{Mean performance of each method over all 21 arrays, with metrics defined in Sec.~\ref{ssec:methodology}. The BSM Noisy row is the noisy binaural signal, unprocessed, as computed using BSM; the BSM Clean row is the same but using the clean target signal only; 
%passes each metric's own clean, noise-free target through the fixed BSM filters with no mask (the best any masking method could reach); neither is ranked. 
%$\Delta$C50 is computed on the clean target alone, so it is exactly 0 for Noisy (no array or BSM processing applied) and reflects only the fixed renderer's own footprint for BSM Clean. 
MIMO FT-JNF and MIMO FT-JNF (Fusion) are the baselines of Sec.~\ref{ssec:baselines}, evaluated on the same full dataset as the proposed method's variants. The best value per column, among the ranked methods, is in bold.}
\label{tab:results}
\end{table*}

\subsection{Setup}
\label{ssec:Setup}

Simulated acoustics scenes are generated with a target speaker and five interfering speakers recorded by an array in a reverberant room, simulated using the image method, with varying 
%room sizes, reverberation times, speech signals and array positions,
acoustic conditions as detailed in \cite{ambidropIEEE}. Clean speech signals are drawn from the WSJ0 corpus~\cite{garofolo2007wsj0}.
Twenty simulated arrays of \cite{ambidropIEEE}, together with a simulated model of the Aria glasses array, include linear, planar and volumetric arrays of various configurations, confined to a sphere of radius 0.1m, each with $D=7$ microphones. These are simulated using an Ambisonics representation of the sound field of order $N=20$ and include both free-field and rigid-sphere steering formulations.
%The listener's ears are located at $\pm0.0875\,$m along the interaural axis of a virtual head coinciding with the center of the array. 
The inference dataset comprises 100 examples per array, 
%with the scenes held identical across arrays for a direct comparison, 
and is disjoint from the training and validation sets.

\subsection{Methodology and Metrics}
\label{ssec:methodology}

In the realization of the proposed method and its variants, the network of Sec.~\ref{ssec:ambidrop} is used unmodified, composed of a two-layer 64-unit BLSTM with a complex-mask output head, 142{,}594 parameters and channel dropout of up to three channels at probability $0.4$, with all parameters identical to \cite{ambidropIEEE}.
For these methods, microphone signals are preprocessed with an STFT using a $32\,$ms Hamming window and 50\% overlap, and encoded by microphone-aligned ASM (Sec.~\ref{ssec:bambidrop}) as a Tikhonov-regularized least-squares solution (Eq.~\ref{eq:c_opt}), with a regularization per frequency as in \cite{madmoni2020beamforming}. \emph{Fusion} is evaluated over the $D$ microphone positions, Eq. (\ref{eq:fusion}).
% ; Sec.~\ref{ssec:results} additionally reports how much a larger center set changes the result. 
The BSM filters (Sec.~\ref{ssec:bambidrop}) follow the same regularized solution (Eq.~\ref{eq:c_opt}), with the magnitude least-squares crossover frequency set as in \cite{madmoni2020beamforming}. These filters are designed once per array and reused unchanged for each of the methods.

The two baselines of Sec.~\ref{ssec:baselines} share the network hyperparameters above and the same negative SI-SDR objective as in \cite{ambidropIEEE}, with one scaling shared across all $D$ microphones. They are trained with the 10 training arrays of \cite{ambidropIEEE} and evaluated on those as well as on the 10 test arrays together with the simulated Aria array (Sec.~\ref{ssec:results}).

The reference binaural signals are rendered from the full $N=20$ order Ambisonics representation of the scene using the Neumann KU 100 HRTF~\cite{bernschutz2013kemar}.
%, and are therefore independent of the order-2 ASM and BSM chain that the evaluated configurations pass through.
% Three references are used: the true direct ears, the true reverberant ears, and the reverberant target split at $50\,$ms after the direct arrival into an early and a late part.
Two such references are used: the \emph{reference binaural signal}, rendered from the full reverberant scene, and the \emph{direct binaural signal}, rendered from the direct-path component alone.
Performance is reported by the following metrics. \textbf{SI-SDR} is measured by the SI-SDR \cite{le2019sdr} of the binaural signals, with respect to the reference binaural signal, averaged over both ears, as are \textbf{PESQ}~\cite{rix2001pesq} and \textbf{STOI}~\cite{taal2011stoi}.
%\textbf{$\Delta$Noise} is measured as the ratio of the interference and noise signals before and after masking, obtained by applying the mask to the signals with the target speaker removed. 
\textbf{SI-SDR-Dir} measures the distortion caused by the mask to the direct sound from the sources, and is the SI-SDR of the masked direct-path signal, with respect to the direct binaural signal, averaged over both ears. \textbf{$\Delta$C50} is measured by the absolute error of the clarity index $C_{50}$~\cite{iso3382} of the masked reverberant target signal with respect to that of the reference binaural signal.
%$C_{50}=10\log_{10}(\mathcal{E}_{\mathit{early}}/\mathcal{E}_{\mathit{late}})$, where $\mathcal{E}$ denotes signal energy.
%
% Binaural cues are evaluated on the enhanced binaural signal against the direct binaural signal.
%since the interaural cues are carried primarily by the direct sound. These measure include 
% {Interaural time difference (ITD) is estimated from the cross-correlation of the binaural signals, low-pass filtered at $1.5\,$kHz ~\cite{faller2004source}, with \textbf{$\Delta$ITD} as the error relative to the direct binaural signal.
%, and the covariance-domain measures defined below.
%
%Two covariance-domain measures are used. 
% computed from $\mathbf{R}(f)=\mathbb{E}_t[\mathbf{p}(t,f)\mathbf{p}^H(t,f)]$, the $2\times2$ spatial covariance of the binaural STFT vector $\mathbf{p}(t,f)$ with the expectation taken over time frames, and include inter-aural level difference, $\mathrm{ILD_{cov}}=10\log_{10}(R_{\mathit{LL}}/R_{\mathit{RR}})$~\cite{vilkamo2017time}, and interaural coherence $\mathrm{IC}=\mathit{Re}\{R_{\mathit{LR}}\}/\sqrt{R_{\mathit{LL}}R_{\mathit{RR}}}$~\cite{bradley1995objective}.
Interaural level difference (ILD)~\cite{vilkamo2017time} and interaural coherence (IC)~\cite{bradley1995objective} are measured from the binaural covariances of the enhanced binaural signal, and evaluated against the same measures computed from the direct binaural signal and the reference binaural signal, respectively, leading to \textbf{$\Delta$ILD} and \textbf{$\Delta$IC}. They are
%$\mathrm{IC}$~\cite{bradley1995objective} measures interaural coherence relative to the reference binaural signal.}
%, which describe lateralization and envelopment in complex sound fields. 
computed as the RMSE over frequency after time alignment.
%with respect to the direct binaural signal, computed 
%A reference row accompanies the results: the BSM ceiling, the unmasked render, which bounds what the fixed renderer can reproduce.

\subsection{Results}
\label{ssec:results}

Table \ref{tab:results} reports the mean over the 100 examples of each array. \emph{Fusion} attains the highest SI-SDR, PESQ, STOI and SI-SDR-Dir, and the lowest $\Delta$IC.
This shows that \emph{Fusion}, by averaging the \emph{Per-Mic} mask estimates and using a shared mask, reduces binaural errors, as it leaves the inter-channel relations that BSM exploits in its design intact, so both enhancement and interaural cues improve. \emph{Center} with its single mask falls only shortly behind, showing that averaging the \emph{Per-Mic} masks is slightly better than using one central mask. 
%\emph{Fusion} is defined over the $D$ microphone positions alone (Sec.~\ref{ssec:bambidrop}); 
An additional study not presented in the table investigated the addition of three more masks to the averaging of \emph{Fusion}, which are centered at the two ears and the center of the head, but this improved the SI-SDR by at most $0.1\,$dB when averaged over all arrays, and so these positions were eventually  excluded from \emph{Fusion}.
%so the microphone positions alone account for the benefit, at no cost of a head model.

\emph{Per-Mic} behaves as anticipated in Sec.~\ref{ssec:altconfig}, with its $\Delta$ILD, SI-SDR-Dir, and $\Delta$IC degrade compared to both \emph{Fusion} and \emph{Center}. In particular, SI-SDR-Dir degrades to $-3.23\,$dB, probably because the independent per-microphone masks modify the channels by different complex gains and break the relations BSM assumed in its design. It nonetheless attains the lowest $\Delta C_{50}$, although the differences to the other methods are not large (about $0.5\,$dB). 
%not from better reverberation preservation, but the same distortion: every configuration over-clarifies the target relative to the reference binaural signal, and the extra late-arriving distortion from independent masking moves the clarity back toward the reference.

%\emph{Fusion} is defined over the $D$ microphone positions alone (Sec.~\ref{ssec:bambidrop}); extending the set to the two ears and array center changes $\Delta$SI-SDR by at most $0.1\,$dB averaged over all arrays, so the microphone positions alone account for the benefit, at no cost of a head model.

% \begin{table}[t]
% \centering
% \begin{tabular}{lcc}
% \hline
% Method & Enh $\Delta$ dB$\uparrow$ & ITD $\mu$s$\downarrow$ \\
% \hline
% BAmbiDrop (Fusion) & \textbf{4.7} / \textbf{5.5} & 66 / \textbf{81} \\
% MIMO FT-JNF & 4.0 / -5.2 & \textbf{60} / 222 \\
% MIMO FT-JNF (Fusion) & 4.2 / -7.2 & 65 / 249 \\
% \hline
% \end{tabular}
% \caption{Fusion vs.\ the two baselines of Sec.~\ref{ssec:baselines}, on their training arrays and the arrays held out from training (each cell: train / test), averaged over all arrays. Best value per column in bold.}
% \label{tab:traintest}
% \end{table}

Table~\ref{tab:results} also includes the two baselines of Sec.~\ref{ssec:baselines}, trained on only 10 of the 21 arrays (Sec. \ref{ssec:methodology}) but evaluated here, like the proposed method's variants, on the full 21-array set. Both collapse well below \emph{Fusion} on every column, several going negative on SI-SDR and SI-SDR-Dir: fitting a mask directly to the microphone domain does not generalize the way the array-agnostic Ambisonics-domain design does, as was already shown in \cite{tatarjitzky2026ambidrop}.

Note that although the values of SI-SDR are not very high for all methods, they are close to that of BSM Clean, which incorporates both residual noise and distortion of the binaural signal, and indicate a high standing reference point. In conclusion, the proposed \emph{Fusion} showed the feasibility of extending the previously proposed AmbiDrop method for array-agnostic speech enhancement to binaural reproduction without any additional training - simply by applying the trained network to ASM computed around each microphone position. 

\section{Conclusions}
\label{sec:conclusions}

BAmbiDrop extends array-agnostic single-channel speech enhancement to binaural reproduction. Relocating the Ambisonics origin allows a frozen AmbiDrop network, trained only on ideal Ambisonics, to provide a mask at any point in space; fusing the masks estimated across the array into one shared complex mask and rendering the masked microphone signals with fixed BSM filters then yields a binaural output with no component retrained. This configuration improved both enhancement and interaural cue reproduction, whereas masking each microphone independently degraded the binaural signals. 
%The target speaker was located in front of the array throughout, although the movable-center construction makes no such assumption.
Future work is proposed to address an off-axis and dynamic target speaker, improving the practical aspects of the method.%by exploiting the rotational properties of the Ambisonics representation, steering the encoded field toward the direction the network expects without re-encoding or retraining.

%\vfill
%\pagebreak

\bibliographystyle{IEEEbib}
\bibliography{BAmbiDrop}

\end{document}